\documentclass[preprint,authoryear,12pt]{elsarticle}

\usepackage{epsfig}

\usepackage {rotating}

\usepackage{amssymb}

\usepackage[
a4paper=true,%
breaklinks=true,%
colorlinks=true,%
pdfauthor={First Author et al.},%
pdftitle={Template for manuscripts in Advances in Space Research}%
]{hyperref}

\journal{Advances in Space Research}

\begin{document}

\begin{frontmatter}



\title{Low ionization lines in high luminosity quasars: The calcium triplet}
\author{Mary Loli Mart\'{\i}nez-Aldama \corref{cor}}
\cortext[cor]{Corresponding author}
\ead{maldama@astro.unam.mx}
\author{Deborah Dultzin}
\address{Instituto de Astronom\'{\i}a, Universidad Nacional Aut\'{o}noma de M\'{e}xico, Aptdo. Postal 70-264, M\'{e}xico, D. F. 04510, Mexico.

}

\author{Paola Marziani}
\address{INAF, Osservatorio Astronomico di Padova, 
Vicolo dell’ Osservatorio 5, 35122 Padova, Italy}

\author{Jack W. Sulentic} 
\address{Instituto de Astronom\'{\i}a de Andaluc\'{\i}a (CSIC),  C/ Camino Bajo de Hu\'{e}tor 50, 18008 Granada, Spain.}\author{Yang Chen}\author{ Alessandro Bressan} 
\address{Scuola Internazionale Superiore 
di Studi Avanzati (SISSA), via Bonomea 265, I-34136 Trieste, Italy}
\author{Giovanna M. Stirpe}
\address{INAF, Osservatorio Astronomico di Bologna, via Ranzani 1, 40127 Bologna, Italy.}
\begin{abstract}

In order to investigate where and how low ionization lines are emitted in quasars we are studying a new collection of spectra of the CaII triplet at $\lambda$8498,  
$\lambda$8542, $\lambda$8662 observed with the Very Large Telescope (VLT) using the Infrared Spectrometer And Array Camera (ISAAC). Our sample involves luminous quasars at  
intermediate redshift for which CaII observations are almost nonexistent. We fit the CaII triplet and the OI $\lambda$8446 line using the H$\beta$ profile as a model. We 
derive constraints on the line emitting region from the relative strength of the CaII triplet, OI $\lambda$8446 and H$\beta$.

\end{abstract}

\begin{keyword}
Line formation; Quasars: general;  Quasars: emission lines
\end{keyword}

\end{frontmatter}

\parindent=0.5 cm

\section{Introduction}

Explaining the origin of Fe emission in quasars is a long-standing problem in active galactic nuclei (AGN) research. The extreme complexity of the FeII ion makes theoretical model calculations very difficult and line blending makes estimation of FeII width and strength parameters uncertain. The Ca$^+$ ion is, by contrast, far simpler. The ionization potential of neutral Calcium is 6.1 eV so we expect CaII ions to exist wherever hydrogen is not fully ionized. Several lines of evidence suggest that CaII IR triplet at $\lambda$8498, $\lambda$8542, $\lambda$8662 (hereafter CaII for brevity) and optical FeII are produced in the same region. Data from \citet{PER88} and photoionization calculations  from  \citet{JOL89} found that CaII is emitted by gas with low temperature (8000 K), high density ($>$10$^{11}$ cm$^{-3}$) and a high column density ($>$10$^{23}$ cm$^{-2}$) similar to the optical FeII. \citet{FERPER89} improved the photoionization models including physical processes like H${^0}$ free-free, H${^-}$ 
bound-free and Compton recoil ionization, that showed the need for very large column  densities ($>$10$^{24.5}$ cm$^{-2}$) to reproduce the CaII spectrum. Such large column density gas could be provided by an accretion disk. A similar behavior for CaII/H$\beta$ and FeII/H$\beta$\ was found by \citet{DUL99}. These authors also suggested that  the line emitting region could be associated with the outer part of an accretion disk. Recently \citet{MAT07} performed  photoionization models and found that low ionization parameters (U $\sim$ 10$^{-2.5}$) are needed to reproduce the flux ratios. Therefore, there is evidence pointing toward low ionization lines like FeII and CaII arising in a region probably associated  with an accretion disk, with physical conditions different from the regions emitting most of the high ionization lines.\\

The present work extends the study of CaII $\lambda$8498, $\lambda$8542, $\lambda$8662 to 14 high luminosity and intermediate redshift quasars that are analyzed within the 4D Eigenvector 1 parameter space context \citep[4DE1;][]{SMD2000a, SUL2000, MAR01, MAR03a, MAR03b, SUL07},  which serves as a spectroscopic unifier/discriminator  of the emission lines properties for type 1 AGN. 4DE1 includes (1) FWHM(H$\beta_{BC}$), (2) equivalent width (EW) of the optical Fe II $\lambda$ 4570 blend and H$\beta$ defined as the ratio R$_{FeII}$= W(Fe II $\lambda$4570)/W(H$\beta$), (3) the soft X-ray photon index ($\Gamma_{soft}$) and (4) the centroid line shift of high-ionization C IV $\lambda$1549 with the strongest correlations involving parameters 2, 3 and 4  \citep{SUL07}. \\

In the 4DE1 parameter framework the broad line AGNs can be divided in two populations, A and B \citep{SUL02}. Considering the broad component of the H$\beta$ line, population A and B can be separated at FWHM(H$\beta_{BC}$) = 4000 km s$^{-1}$; spectra of sources above and below this limit look substantially different. Population A shows: (1) a scarcity of RL sources, (2) strong/moderate Fe II emission, (3) a soft X-ray excess, (4) high-ionization broad lines with blueshift/asymmetry and (5) low-ionization broad line profiles (LIL) best described by Lorentz fits. Meanwhile, Population B: (1) includes the large majority of the RL sources, (2) shows weak/moderate Fe II emission, includes sources (3)   with less prominent or no soft X-ray excess   \citep{SUL07} and (4) with HIL blueshift/asymmetry or  no blue shifts at all. Last, (5) Pop. B shows LIL Balmer lines best fit with double Gaussian models. The physical drivers that  change along the 4DE1 sequence have been identified: number density appears to increase 
from Pop. B to A, 
and black hole mass increases with large scatter from A to B. The principal driver of source occupation in 4DE1 space involves Eddington ratio which increases from B to A \citep{MAR01}. \\

This paper presents new observations and data reduction of spectra of the CaII IR triplet (\S \ref{obs}), as a representative low ionization lines, which are used to try to know the physical conditions from the emitting region. We discuss in some detail  the line and continuum components identified for a proper data analysis (\S \ref{an}). Preliminary results obtained for the comparison between observational data and photoionization predictions are briefly reported in \S \ref{prelim}. Conclusions are present in \S \ref{end}.

\section{Observations and data reduction }
\label{obs}

\subsection{Sample selection and observations}
Up to recent time CaII samples were selected in sources with strong FeII emission and low redshift \citep{PER88, MAT07}. CaII had been seen in absorption, coming from the host galaxy of the AGN, so that a strong FeII emission guarantees the observation of  CaII in emission from  the  broad line region (BLR). Our sample was selected without considering FeII intensity. We selected sources where CaII and OI lines were not affected by  atmospheric absorption present in the infrared region. Also, we only chose targets where the optical spectrum around H$\beta$ has already been observed.\\

Following the 4DE1 context, population A sources show strong FeII emission and therefore we expect CaII emission to be strong. The majority of previous observations involved population A sources where both ions show strong emission. Our sample  contains 4 Pop. A and 10 B sources making lower intensity FeII sources well represented in our sample. 
Our sample therefore explores a domain in FeII emission where the study of CaII could give new hints about the origin of low ionization lines.\\

The sample contains 14 high luminosity Hamburg-ESO quasars with a spectral resolution, R$_{S}$, $\sim$ 300, M$_B$ $<$ -26 and 0.847 $< z <$ 1.638 observed at the Very Large Telescope (VLT) equipped with the  Infrared Spectrometer And Array Camera (ISAAC) during 2010, in service mode.  Table 1 lists  target name, redshift, absolute magnitude, Kellermann's coefficient for radio-loudness  and spectral type in the Eigenvector 1 sequence. The last column reports the standard star name used for sensitivity function calibration.\\

VLT is made up by four telescopes of 8.2 m  diameter located on top of Cerro Paranal. Each telescope operates with a large collection of high  quality instruments. ISAAC is one of them, and has been for many years one of the few instruments available worldwide for moderate resolution IR spectroscopy of faint sources like high-$z$ quasars.  It is a camera able to obtain images and spectra with a high resolution at 1 - 5 $\mu$m  \citep{MOO98} and it is equipped with gratings for high (R${_S}$ $\sim$ 3000) and low resolution (R${_S}$ $\sim$ 500) for a 1'' slit width. Our spectra were collected at low resolution but with a slit width of $0.6$'' that ensured a spectral resolution R${_S} \approx$ 1000.\\

\subsection{CaII - data reduction}

Data reductions were performed using IRAF software. Sequences of frames were obtained alternating the source placement at different locations (e.g. A, B) along the slit.  All frames with the same source location were averaged. The average  observation  at one location was subtracted from the one at the different position to obtain two background subtracted frames (e. g. \={A} - \={B}, and \={B} - \={A}). The resulting differences were divided by the  appropriate flat field frames. Spectra were extracted using the IRAF program \texttt{ apsum} and were calibrated on wavelength with  a xenon/argon arc spectrum that was extracted from the calibration frame. The wavelength scale was set using 3rd order Chebyshev polynomial fits to the positions of the most intense lines in the H and K band,  1.5 - 2 $\mu$m and 2 - 3 $\mu$m respectively. Once matched with the  corresponding arc calibrations, spectra of each source were rebinned to a linear wavelength scale whose zero point was readjusted  using   suitable skylines.
 Frames  were then averaged with weights proportional to their total integration  time.\\

The spectra 
of the telluric standard stars were extracted and wa\-ve\-length-calibrated in the same way. We eliminated the atmospheric features by dividing the quasar and the standard star spectra by a synthetic atmospheric transmission spectrum. The majority of the standard stars are spectral type B. Unfortunately, we could not find   theoretical  libraries of spectra with a consistent model for reproducing the spectrum in the H and K band. Therefore we decided to use a black-body model corresponding to the temperature of the star determined on the basis of its tabulated spectral type, since this approach yielded  a good approximation to stars with observed IR spectral energy distribution. The sensitivity function was  then obtained   dividing the standard star spectrum by the black-body model. \\

Finally, the correct flux calibration was achieved by scaling the standard star spectra according to its magnitude. Since the seeing almost always exceeded the width of the slit, a significant light loss occurred and an additional correction had to be applied. Although extinction is not large in this region, the spectra were corrected using the data given by \citet{SFD98}. Redshift correction was carried out assuming redshift values reported by \citet{SUL04} and \citet{MAR09} based on H$\beta$$_{NC}$, 
H$\gamma _{NC}$ and  [OIII] $\lambda$$\lambda$5007,4959 with an uncertainty usually $<$150 km s$^{-1}$. The right-hand panels of Figure 1  show examples of two calibrated CaII spectra.\\

\subsection{H$\beta$ observations}
CaII observations are complemented with optical spectra that cover all or part of the H$\beta$, FeII $\lambda$4570 and/or FeII  $\lambda$5130 blend. Optical measurements were obtained with VLT ISAAC in service mode between 2001 and 2005 \citep{SUL04, MAR09}. Since H$\beta$ and CaII spectra were observed at widely different epochs ($\approx$ 10 yr apart), both spectra were normalized to 2MASS magnitudes, setting a normalized flux scale of the same epoch to avoid variability effects (the response times of the BLR to continuum changes is several years). HE0048-2804 and HE2340-4443 were observed in the Z band to cover H$\beta$. 2MASS data are not available in Z, so that no normalization was applied to their H$\beta$ spectra. The flux  scale is therefore the one obtained from the standard star. Two examples of H$\beta$ are shown in the left  panels of Figure 1.\\

\section{Data analysis}
\label{an}
We built models of the spectral regions using \texttt{specfit}, an IRAF task that preforms simultaneous fits of various line and continuum components \citep{KRI94}. Each component has  a specific number of associated parameters. Providing a lower and an upper limit, and an initial guess value for each parameter, \texttt{specfit} can compute a minimum $\chi^2$ \ fit.  At first we performed the H$\beta$ fits, allowing all   parameters to vary freely. Although \citet{SUL04} and \citet{MAR09} provide previous ones for H$\beta$, we decided to redo all measurements for H$\beta$ as aid for processing high order Paschen lines, where the same FWHM and shift were fitted. For CaII and OI we only took the H$\beta$ FWHM as an initial guess, leaving  \texttt{specfit} free to compute  the best model within reasonable upper and lower limits. Two examples of  the fits are shown in Figure 2 (left: H$\beta$, right: CaII). In the following we  describe the component used for the fits.\\

\begin{enumerate}
\item\textit{Continuum}. We assume that the continuum underlying optical and near-infrared regions is a power law with variable slope. For H$\beta$ it was fixed taking as reference 4750 and 5100 \AA. Due to the small bandwidth of our IR spectra, the identification of the continuum was difficult. Depending on the wavelength range in each spectrum, we fit a local continuum taking one or two reference points at 8100, 8800 and 9400 \AA, obtaining a different slope for each case.

\item\textit{FeII template}. For the FeII optical contribution in H$\beta$ we used the template previously employed by \citet{MAR09}. In the past, the main  efforts in 
modelling the  FeII contribution were focussed on the ultraviolet and optical regions. Only recently \citet{GAR12} produced the first semi-empirical near-infrared template based on the I Zw 1 spectrum.  We used the template kindly made available to us by \citet{GAR12} to model  the FeII emission. The main FeII contributions are  at 9200 \AA\  close to Pa9 $\lambda$9229. According to Garcia-Rissmann, in I Zw 1 the contribution is $\sim$50$\%$\ of Pa9. This relative intensity is however not always observed since there are objects in which  FeII is visually stronger as,  for example, HE0248--3628. We also found a slight FeII deficit in the blue side of Pa9. In those cases we preferred to use the theoretical values reported in the paper  by \citet{GAR12}, and found that the fits improved. In several spectra we observe a little bump at 8200-8300 \AA\ that cannot be reproduced using either the theoretical or the semi-empirical template. 

\item\textit{Broad component.}  Following 4DE1 we assume that low ionization lines in Pop. A and B sources (FWHM H$\beta_{BC}<$ and $>$4000 km s$^{-1}$\ respectively)  
show Lorentzian and double Gaussian profiles in the broad component. Therefore all the broad components were fitted using a Lorentzian in Pop. A sources and a Gaussian in Pop. B sources. The H$\beta$ line was taken as a reference to carry out the fits of the CaII blend, since its profile is well defined. The H$\beta$ FWHM and shifts were also applied to high order Paschen lines present in the near-infrared spectra. Even if H$\beta$ and OI are emitted in a region with similar physical conditions (see below), the FWHM is not the same: FWHM(OI$_{BC}$) $\leq$ FWHM(H$\beta$$_{BC}$), although OI FWHM was allowed to vary over a wide range. CaII triplet was modelled using  three BCs of equal intensity  and with the same FWHM, since the CaII triplet is optically thick \citep{FERPER89}. As CaII lines are completely blended it is difficult to measure a shift, and so the CaII lines were held fixed at rest-frame wavelength.

\item\textit{Very broad component.} A red asymmetry in population B sources with H$\beta$ \ FWHM $\gtrsim$ 4000 km s$^{-1}$\ has been associated with the existence of a distinct emitting region, the so-called Very Broad Line Region \citep[VBLR;][]{MAR09}. A VBC was fit to H$\beta$, OI$ \lambda$8446 and the most intense high order Paschen lines: Pa9, Pa8 and Pa7. The shift and FWHM were assumed  equal in H$\beta$ and Paschen lines. Simulation results indicate  that OI and H$\beta$ are emitted in similar  regions, whose physical conditions are less restrictive than those needed for significant CaII emission (see \S \ref{emission}). Therefore a VBC was not fitted to  CaII emission.  
 
\item\textit{Narrow component.} The narrow lines or components present in the optical spectra are [OIII] $\lambda \lambda$5007, 4959 and H$\beta$. Due to the low S/N in near-infrared spectra it is usually not possible to detect  narrow line emission (that is anyway weak in most cases): only in HE1349+0007 we were able to fit the NC of OI. We could also detect  [SIII] $\lambda$9531. 

\item\textit{High order Paschen lines.} The most intense Paschen lines   in our spectra are Pa9, Pa8 and Pa7. However, if we  only considered these lines  along with  FeII, we would get a deficit of emission at 8700-9100 \AA, so we have decided to include high order Paschen lines. Because Pa9 is present in all our spectra, we decided to take it as a reference. Higher order Paschen lines were scaled using the results of CLOUDY simulations (version 08.00) \citep{FER98}. We noted the presence of the high order Paschen lines form a pseudo-continuum  that cannot be  neglected. This pseudo-continuum covers part of the CaII triplet region specially at  $\lambda$8662. Its inclusion has an effect on the 
intensity and FWHM of the CaII triplet. Given that our  objective is to try to model the emission in this region as accurately as possible, we decided to consider  high order  Paschen lines from Pa9 to Pa24. \citet{PER88} found that Pa14 $\lambda$8598 contributes $\sim$12$\%$ of the flux of  $\lambda$8662 in Mrk 42, so it was ignored. However in HE0035--2853, we found that the contribution of Pa14 is $\sim$70$\%$, while in HE0048--2804 CaII emission is insignificant and Paschen lines dominate the fit. 

\item\textit{Paschen continuum.} Two of our targets, HE0058-3231 and HE2202-2557 cover the head of the  Paschen continuum (PaC) at 8204 \AA. It is interesting to note that previous quasar studies have neglected the PaC due to the expectation that its contribution would be smaller than the Balmer continuum. We could detected a  hint of PaC  in HE0058--3231 and HE2202--2557. In the other sources we could  only cover the region beyond the Paschen limit. A CLOUDY simulation with $\log (U) = -2.5$\ and $\log(n_\mathrm{H}$) = 12 provided us with predictions on the relative intensity of Pa9 and of the integrated PaC \citep[c.f.][]{OF05}. We then estimated the continuum specific intensity at the Paschen edge, and assumed an exponential  decrease   toward shorter wavelengths appropriate for an optically thin continuum. Our estimates appear to be in excess with respect to the observations since the best fits require a PaC smaller than that predicted. There are two main possible explanations: 1) CLOUDY computations 
overpredict the recombination continuum, or 2) the quasar continuum level we set  is not correct; the actual continuum is lower. Unfortunately, we cannot test these options  on our data since the spectra cover only a small wavelength range. To properly define PaC we need to  cover a spectral range spanning from the optical to the NIR. We will try to address the issue of the PaC intensity  in  the future.  

\item\textit{Stellar absorptions.} To subtract the stellar component we used  new stellar population synthesis models (Chen et al. in preparation), based on the code of \citet{BGS98} with updated stellar evolutionary tracks \citep{BRE12} and stellar atmospheres suitable for the analysis of stellar absorption lines in the optical and near infrared spectral regions \citep{SAN06a, SAN06b, SAN06c, RAY09}. We found that the underlying stellar absorption of the host galaxy is significant only  in HE2202-2557, with a luminosity contribution of $\sim$20$\%$, while the rest of the sample was affected only by $<$10$\%$. The stellar population synthesis model was selected  taking into account redshift  (that sets the maximum age of the host galaxy) and typical black hole mass: we assumed a spheroid  mass of 1.13 x10$^{12}$ M$_\odot$, an age of 2.4 Gyr, and   metallicity of 2Z$_\odot$. These properties are consistent with massive ellipticals expected to host very luminous quasars at intermediate redshift.

\end{enumerate}

\section{Preliminary results}
\label{prelim}

\subsection{Observations vs. photoionization models}
\label{emission}

If we consider the broad component only, the flux ratios in our sample are approximately $\log$(CaII/OI) $\approx$ 0.01, $\log$(CaII/H$\beta$) $\approx$ -0.7, $\log$(CaII/Pa9) $\approx$ 0.35, $\log$(OI/H$\beta$) $\approx$ -0.8 and  $\log$ (OI/Pa9) $\approx$ 0.1.  We carried out model calculations in the photoionized BLR using the code CLOUDY version 08.00 \citep{FER98}. The gas was modeled with a column density of 10$^{23}$ cm$^{-2}$  and (U, n$_H$) sets of 7 $ < \log n_H <$13 and -4.5 $< \log U <$ 0. Chemical composition was assumed solar.  The measured ratios,  compared to those predicted by the simulations,  consistently indicate n$_H > 10 ^{11.5}$ cm$^{-3}$ for CaII in agreement  with previous works \citep{JOL89, FERPER89, MAT07,MAT08} and with the physical conditions found for FeII. The ionization parameter is constrained to be $\log U < -1.5$. 

\subsection{Where are OI and CaII emitted?}

OI emission is favored at somewhat lower density (n$_H < 10 ^{11}$ cm$^{-3}$) and higher ionizing photon flux than CaII; in other words, the behaviour of OI is more similar to 
that of H$\beta$. Since H$\beta$ and OI are expected to be emitted in similar regions on the basis of photoionization predictions, we  include a VBC for the OI line of 
population  B sources. This means that emission of H$\beta$\ and OI are assumed to occur in both a Broad Line Region and a Very Broad Line Region associated with the BC and VBC 
respectively. A good fit to the profile blend OI+CaII is possible in this case. The resulting OI  profile is consistent with H$\beta$ BC+VBC and  broader than CaII for which we 
include a BC only.  We stress again that CaII is  expected  to be emitted  mainly in the BLR associated with the BC only, since the ionization  parameter and density in the VBLR are believed to be unsuitable for significant CaII emission.\\ 

Combining our data and Persson's we can see that the original relation between FWHM of CaII, OI $\lambda$8446 and  H$\beta$\ found by \citet{PER88} himself is basically confirmed: FWHM(H$\beta_{BC}$) $\sim$ FWHM(CaII) $\sim$ FWHM(OI$_{BC}$) (Fig. 3). There are however a few Pop. A sources that have a significant broader CaII. Our sources are too low S/N to claim that this difference is real. The data of Persson were analyzed without several emission features (FeII, OI narrow component, high order Paschen lines, ...) that we now know are blended with OI and CaII. To ascertain that FWHM(OI) $<$ FWHM(CaII) in Pop. A, a new analysis of the old Persson's data should be carried out.\\

On the other hand, taking the full profile (BC+VBC) in Pop. B sources for H$\beta$ and OI, we obtained the following average values: FWHM(H$\beta_{BC+VBC}$) = 6100 km s$^{-1}$, FWHM(OI$_{BC+VBC}$) = 5500 km s$^{-1}$ and FWHM(CaII) = 4600 km s$^{-1}$: FWHM(H$\beta_{BC+VBC}$) $>$ FWHM(OI$_{BC+VBC}$) $>$ FWHM(CaII) (Figure 3). This second sequence reflects the likely possibility of an ionization stratification from the innermost zones  where ionization is too high for OI and CaII emission, to the outermost ones where CaII emission if favored. \\

The  equivalent width of CaII is distributed over a wide range, whereas the equivalent width of OI  is not. This result also  indicates that OI is unlikely to be emitted exclusively in the very same region of the CaII triplet. OI is pumped by Bowen fluorescence, but CaII is not and Ly$\beta$ photons can ionize Ca$^+$ to Ca$^{++}$\ from Ca$^+$ ground level. Pumping is especially effective at the boundary between the fully and  partially ionized zone (PIZ), where OI can be neutral. Therefore significant emission is expected whenever the Ly$\beta$ opacity is high and the gas is optically  thick to the ionizing continuum, so that a PIZ can exist. Large column density is necessary. \\

The previous results suggest that OI emission is not restricted to the CaII emitting region. We cautiously considered  VBC emission of OI on the basis of appearance of the OI+CaII blends and of the photoionization simulations prediction, assuming that no OI VBC  implies stronger CaII emission. Our conclusions  will be little affected in this case because, at low ionization parameter, the CaII / Pa$9$\ ratio shows a steep increase with density at $\log n_H > 11 - 11.5$.  Only if the  CaII triplet  were not detected at all, our conclusions would be affected. We plan, however, to study the effect of  different emissivity laws on line profiles in a forthcoming study (Mart\'{i}nez-Aldama et al., in preparation). \\

Both CaII and OI are emitted in the BLR associated with the BC along with other higher ionization lines. We can figure out  a configuration that is plane parallel  with a distribution of   clouds/filaments   above and below the accretion disk. Some of the clouds might be at the same distance from the ionization source as  the dense  CaII emitting regions within the accretion disk. Therefore, clouds and disk may share the same dynamics but  they do not necessarily share the same physical conditions. This provides an explanation for the similarity of the BC profiles; a slightly larger emissivity-weighted distance for OI  could explain  some FWHM differences,  for example FWHM(CaII) $>$ FWHM(OI$_{BC}$).  \\

\section{Conclusions}
\label{end}

We analyzed a new sample of CaII IR triplet and OI $\lambda$8446 in luminous quasars at intermediate redshift in order to investigate  the physical conditions required for low ionization line emission. This work extends previous samples including several quasars with broader lines (Pop. B) and modest or weak FeII emission. For the first time we  include a FeII NIR template, high order Paschen lines and Paschen continuum in the analysis of the spectral region around CaII triplet. According to photoionization predictions, CaII is emitted in a region with n$_H > 10 ^{11.5}$ cm$^{-3}$\ and $\log U < -1.5$. These conditions are  similar to those capable of accounting for the FeII emission in a photoionization scenario. Meanwhile OI emission is favored  n$_H < 10 ^{11}$  cm$^{-3}$ and higher ionizing photon flux. The physical conditions for OI are more similar to H$\beta$, so that we included BC+VBC emission in Pop. B., sources for OI $\lambda$8446. This yielded  a better fit of the whole OI+CaII blend. \\

We thank both reviewers for their useful comments to improve this manuscript. D.D. acknowledges support from Grant IN107313, 
PAPIIT UNAM.


\begin{figure}[ht]
  \begin{center}
      \includegraphics[width=14cm,keepaspectratio=true]{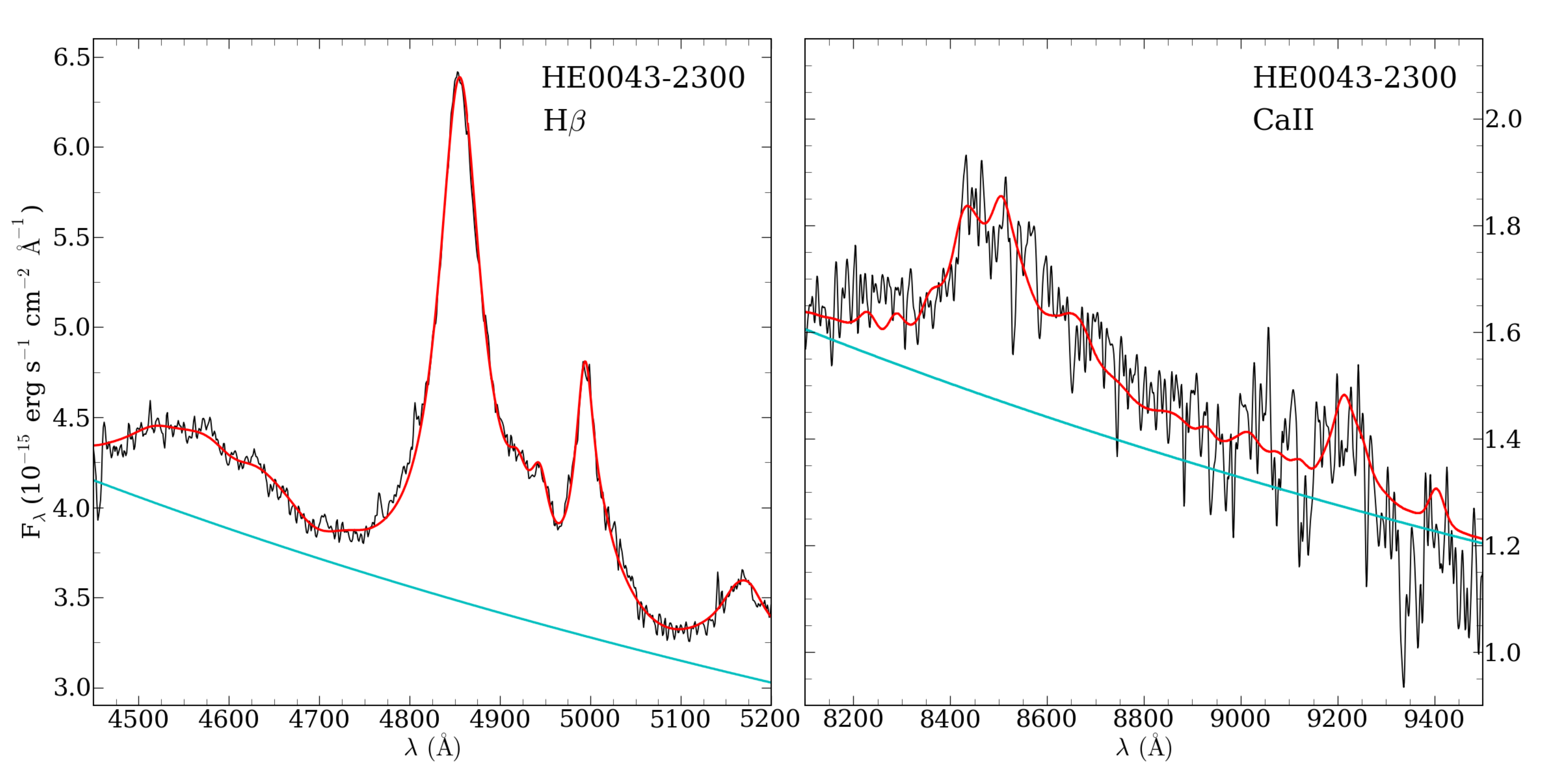}
      \includegraphics[width=14cm,keepaspectratio=true]{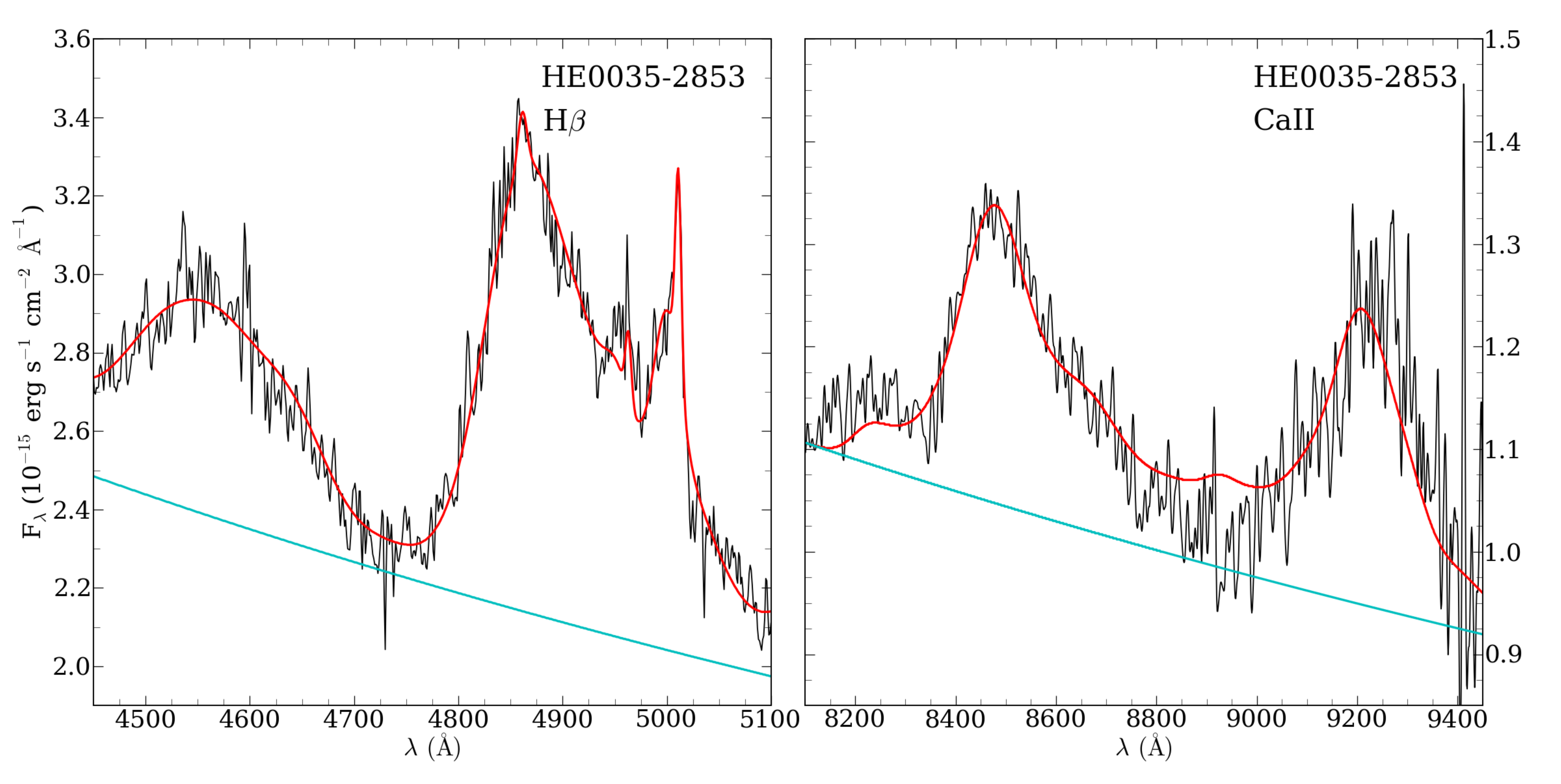} \\
      \caption{Example of intermediate-redshift quasars. The top panel shows a source representative of Pop. A, the bottom one of Pop. B. Abscissae are 
      rest-frame wavelength in \AA, ordinates are specific flux in units of 10$^{-15}$ erg s$^{-1}$ cm$^{-2}$ A$^{-1}$. The left panels show H$\beta$ 
      spectral region before continuum subtraction. Right panels show CaII spectral region. In both panels the best fit is marked by the red color line 
      and the continuum level is marked with the cyan line.}
     \label{hbca}
  \end{center}
\end{figure}

\begin{figure}[ht]
  \begin{center}
      \includegraphics[width=6.5cm,keepaspectratio=true]{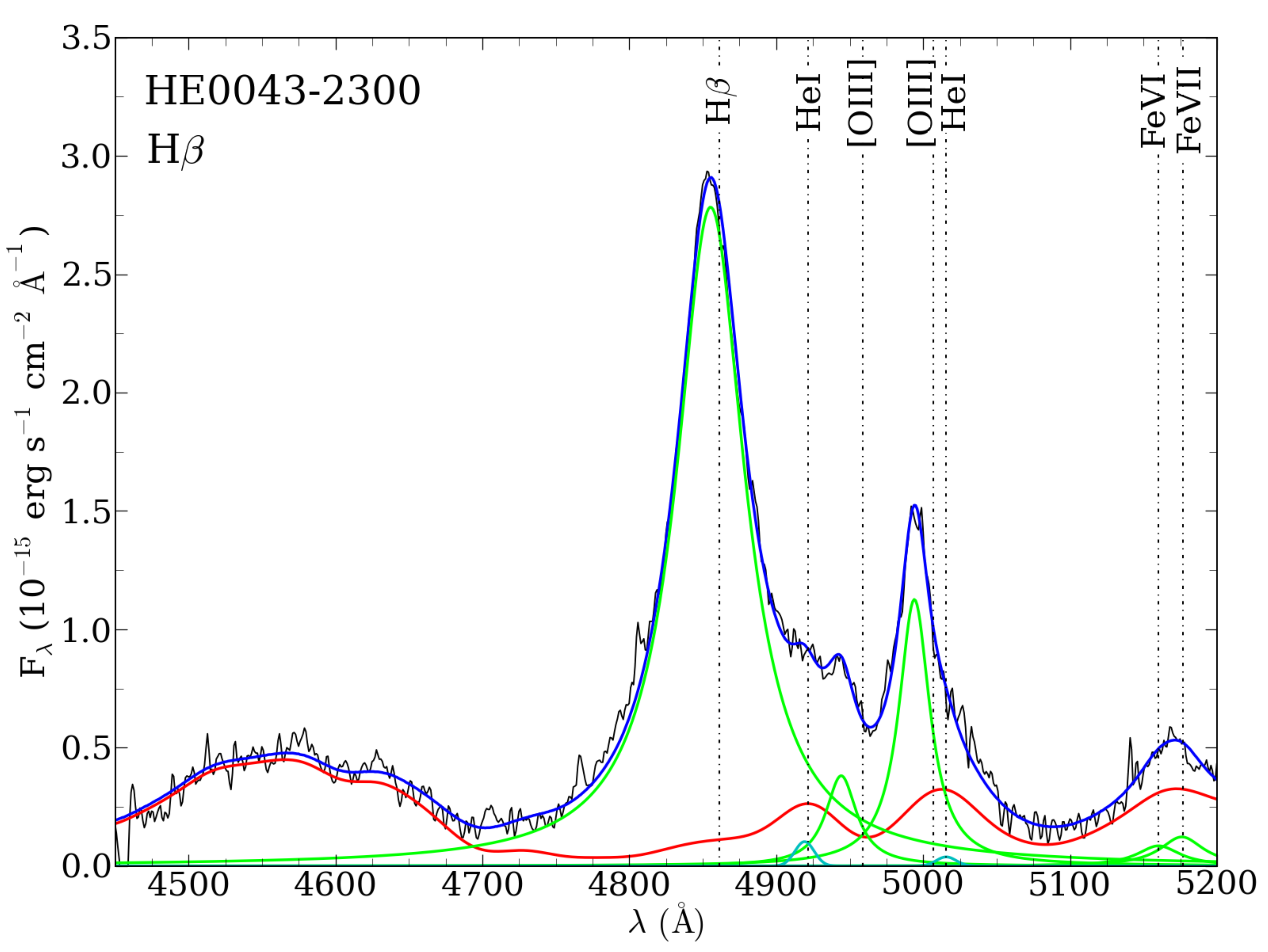}
      \includegraphics[width=6.5cm,keepaspectratio=true]{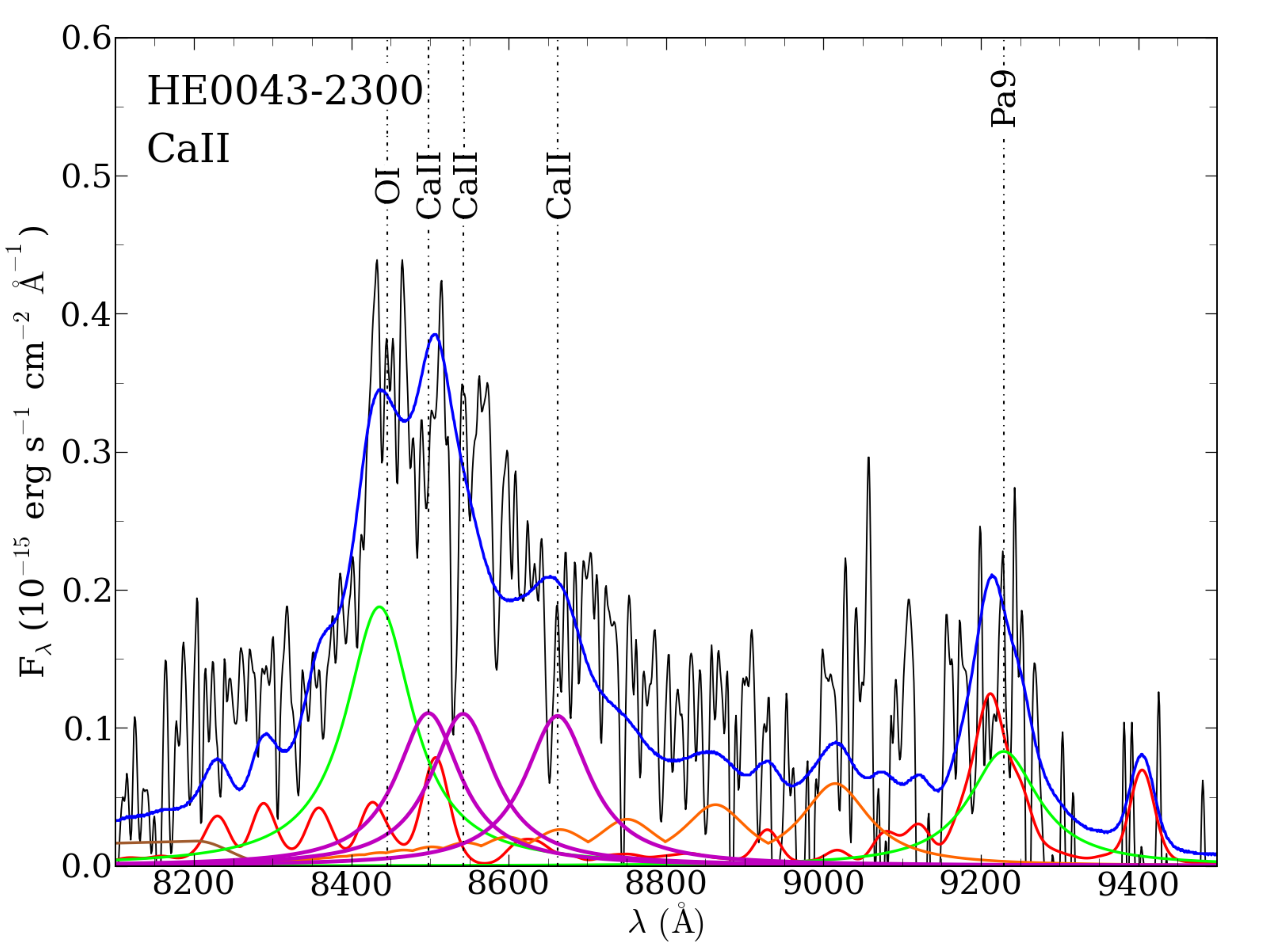} \\
      \includegraphics[width=6.5cm,keepaspectratio=true]{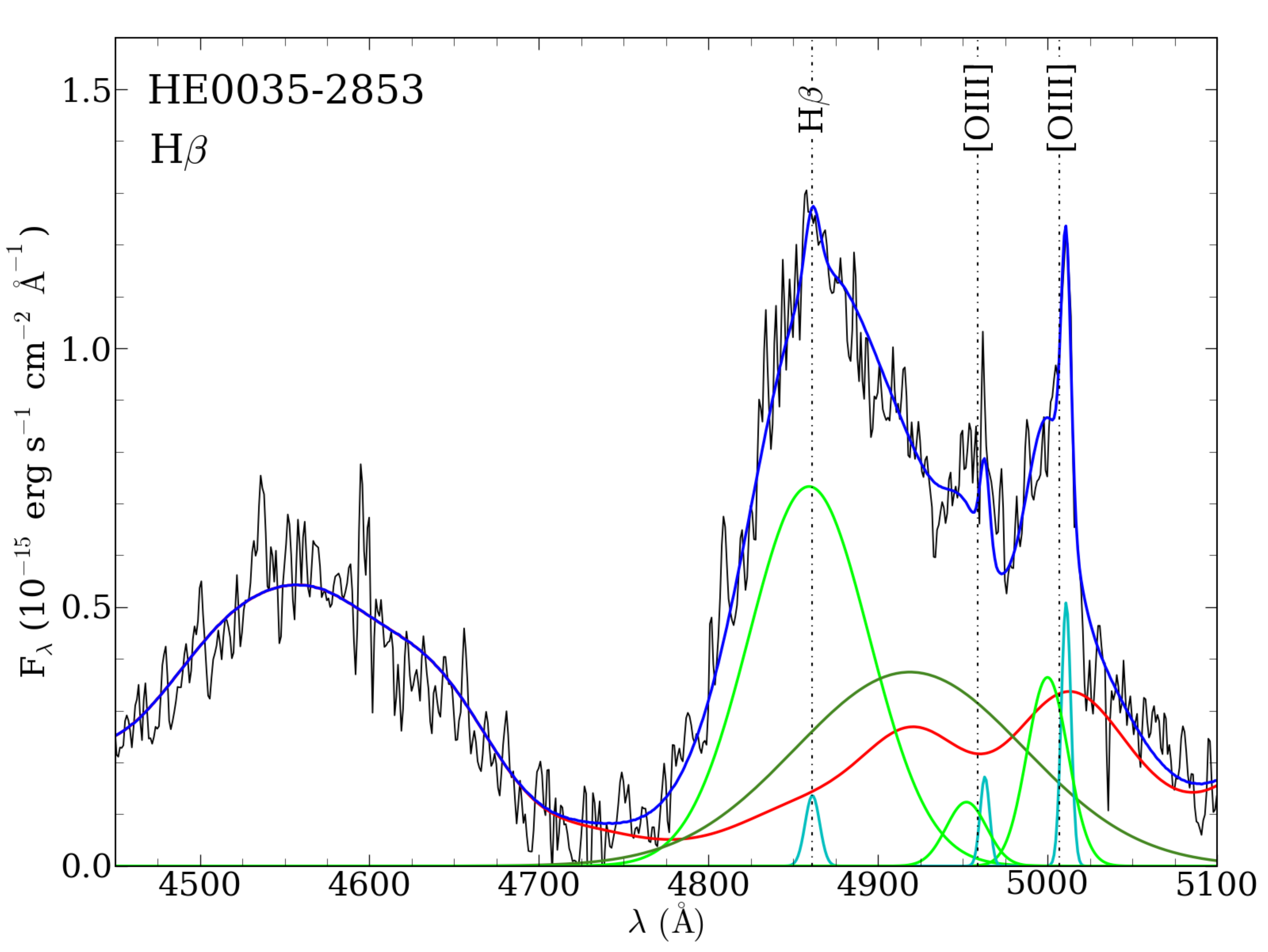}
      \includegraphics[width=6.5cm,keepaspectratio=true]{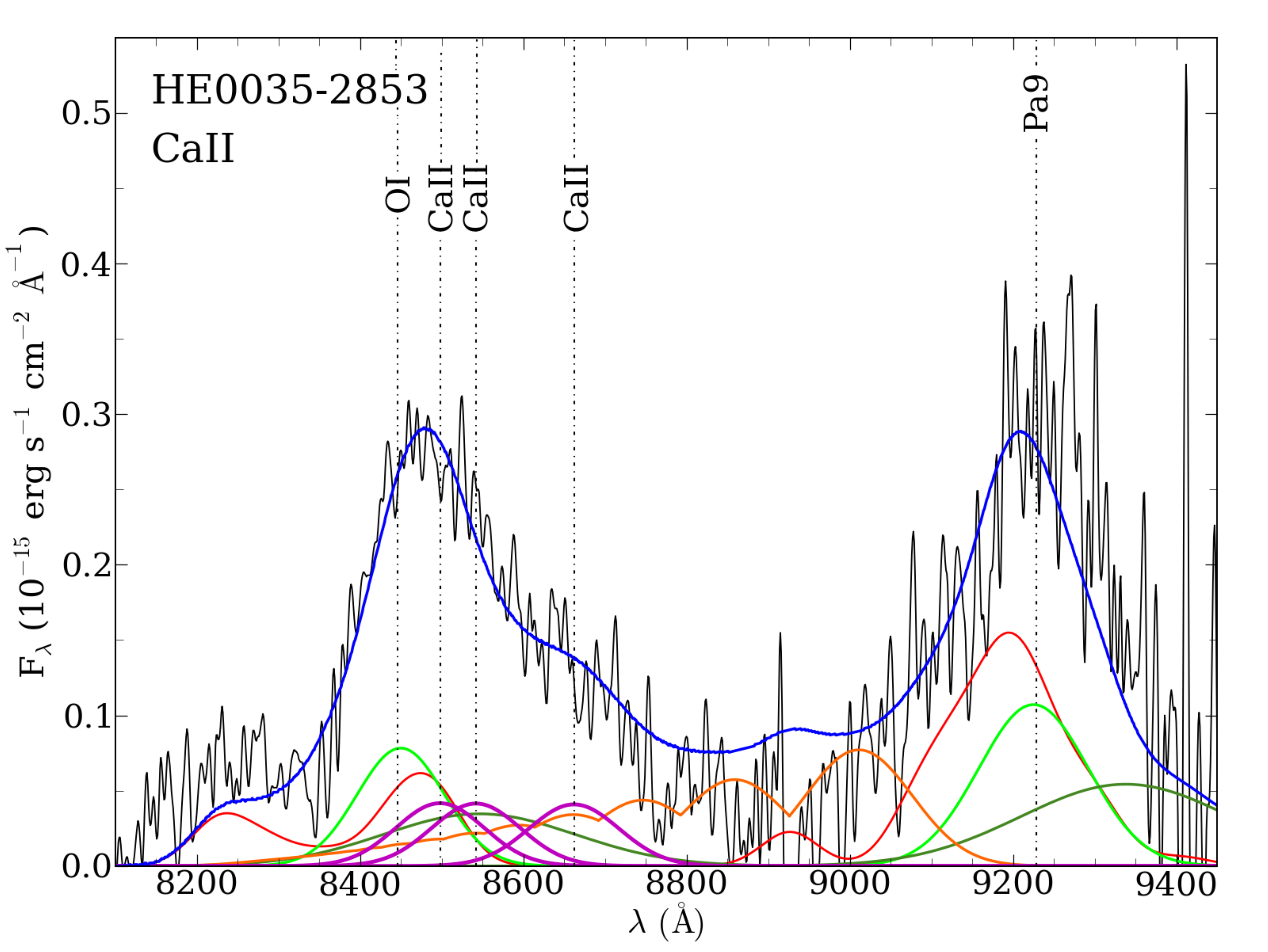} \\
      \caption{Continuum-subtracted spectra for H$\beta$ and CaII spectral regions (left and right panels). Blue lines show the best fit. BC is marked by the 
      fluorescent green, VBC is marked by the green line, NC is marked by the cyan line and the magenta line marks the CaII triplet. Red line traces 
      FeII emission and orange line traces the pseudo-continuum of High Order Paschen Lines. The vertical dash-point lines mark the rest-frame wavelenght of the 
      lines fitted. Abscissae are rest-frame wavelength in \AA, ordinate are specific flux in units of 10$^{-15}$ erg s$^{-1}$ cm$^{-2}$ \AA$^{-1}$. In H$\beta$ 
      spectra can be observed [OIII] $\lambda\lambda$5007, 4959, both spectra present blueshifted broad component commonly asociated to a wind. Only HE0035-2853 
      present a narrow component the rest-frame for H$\beta$ and [OIII]. In HE0043-2300 could be observed a weak emission of HeI $\lambda$4921, HeI $\lambda$5015, 
      FeVI $\lambda$5160 and FeVII $\lambda$5177. In the infrared spectra for both cases, OI and CaII triplet are completely blended, with OI most intense than CaII. In 
      HE0035-2853 high order Paschen lines contribute largerly to the total spectrum.}    
     \label{prof}
  \end{center}
\end{figure}

\begin{figure}[ht]
  \begin{center}
      \includegraphics[width=13cm,keepaspectratio=true]{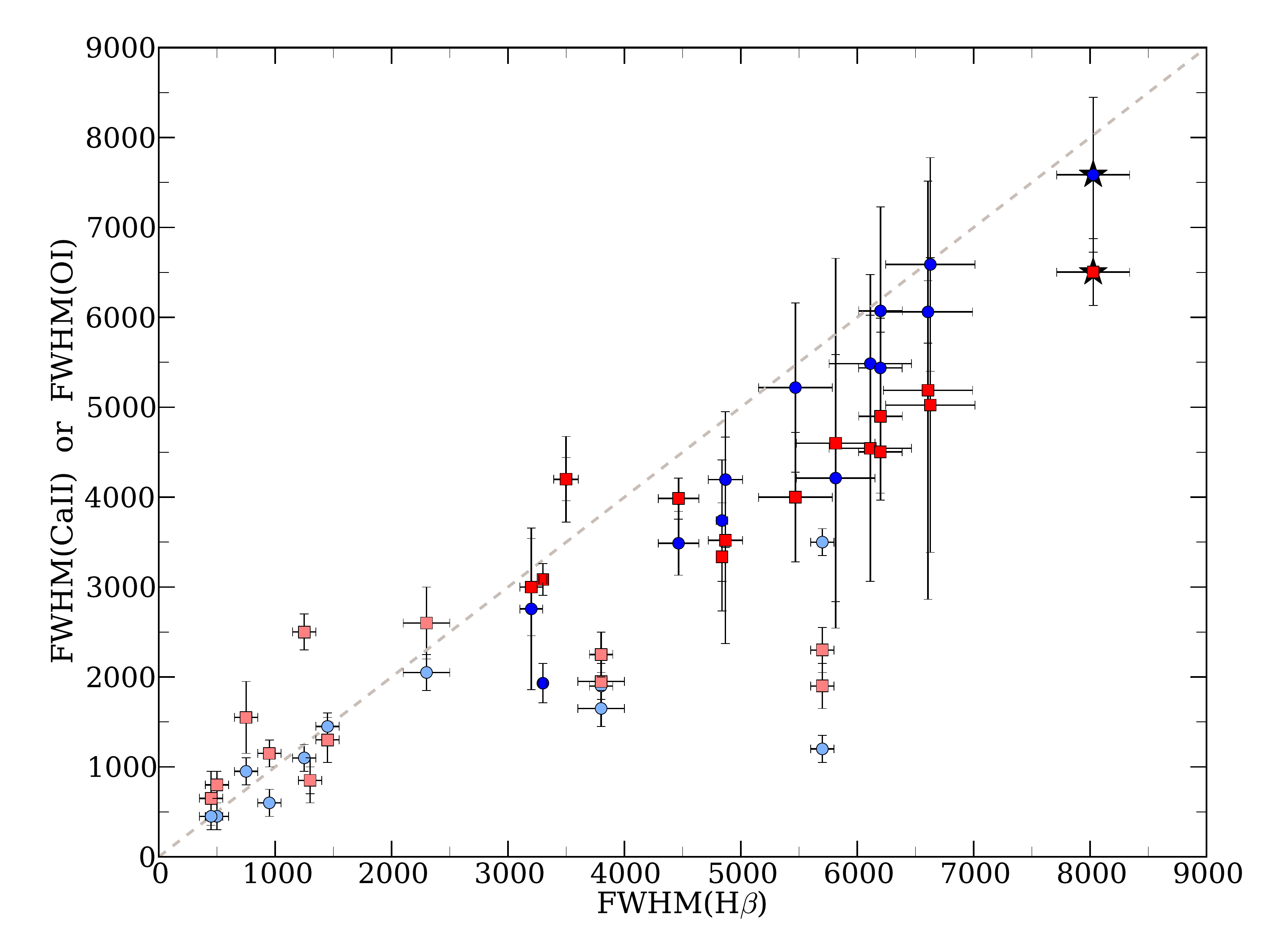}
      \caption{Comparison between CaII (BC), OI (BC+VBC) and H$\beta$ (BC+VBC) FWHMs in km s$^{-1}$. Abscissa is the H$\beta$ FWHM, ordinate is the CaII 
      Triplet and OI $\lambda$8446 FWHM. Blue circles: OI data. Red squares: CaII data; pale color data points belong to the Persson's sample \citep{PER88}. ç
      Starred point is  HE2202-2557, the quasar with significant stellar absorption. The dashed line has a slope of unity.
      }
     \label{comp}
  \end{center}
\end{figure}

\begin{table}[H]
\caption{Basic properties of sources}
\centering
\begin{center}
  \begin{tabular}{cccccc}
  \hline
  Target name &z $^1$ &M${_B}^{2}$ &log(R${_K})^{3}$ &Sp. T.$^{4}$ &Standard Star $^{5}$ \\
  \hline \hline
    HE0005$-$2355	&1.412 &-27.6 &2.56	&B1 &Hip000183\\
    HE0035$-$2853	&1.638 &-28.1 &$<$0.21	&B  &Hip005988\\
    HE0043$-$2300	&1.540 &-27.9 &2.03	&A1 &Hip005988\\
    HE0048$-$2804	&0.847 &-26.0 &$...$	&B1 &Hip005988\\
    HE0058$-$3231	&1.582 &-27.9 &$<$0.24	&B1 &Hip005988\\
    HE0203$-$4627	&1.438 &-27.5 &2.07	&B2 &Hip012248\\
    HE0248$-$3628	&1.536 &-28.2 &0.55	&A1 &Hip108612\\
    HE1349$+$0007	&1.444 &-28.0 &-0.18	&B  &Hip082037\\
    HE1409$+$0101	&1.650 &-28.3 &0.40	&B  &Hip088609\\
    HE2147$-$3212	&1.543 &-28.2 &$<$0.14	&B  &Hip099286\\
    HE2202$-$2557	&1.535 &-28.1 &1.80	&B1 &Hip111563\\
    HE2340$-$4443	&0.922 &-26.3 &$...$	&A1 &Hip111563\\
    HE2349$-$3800	&1.604 &-27.4 &1.93	&B2 &Hip104374\\
    HE2352$-$4010	&1.580 &-28.8 &$...$	&A1 &Hip001904\\
    \hline
  \end{tabular}
\end{center}

\footnotesize \raggedright
$^1$ Redshift, uncertain and reference in \citet{SUL07} $\&$ \citet{MAR09}\\ 
$^2$ Absolute magnitude, reference in \citet{SUL07} $\&$ \citet{MAR09}\\
$^3$ Decimal logarithm of the specific flux at 6 cm and 4440 \AA\ \\
$^4$ Spectral type in the Eigenvector 1 sequence: Population A or B. \\
$^5$ Standard star name used for sensitivity function calibration.\\

\end{table}

\end{document}